\documentclass[acmtog, nonacm]{acmart}

\usepackage{enumitem}
\usepackage{booktabs}
\usepackage{subcaption}

\AtBeginDocument{%
  \fancypagestyle{plain}{%
    \fancyhf{} %
    \fancyfoot[L]{ACM SIGENERGY Energy Informatics Review}%
    \fancyfoot[R]{Volume 6 Issue 2, July 2026}}
  \providecommand\BibTeX{{%
    \normalfont B\kern-0.5em{\scshape i\kern-0.25em b}\kern-0.8em\TeX}}}

\let\oldmaketitle\maketitle
\renewcommand{\maketitle}{%
  \oldmaketitle%
  \thispagestyle{plain}%
  \pagestyle{plain}}

\begin{document}

\title[The Cost of Context: Profiling the Energy Footprint of Input Tokens in Large Language Models]{The Cost of Context: Profiling the Energy Footprint\\of Input Tokens in Large Language Models}
\titlenote{Author's version. Accepted at HotCarbon'26, the 5th Workshop on Sustainable Computer Systems; to appear in ACM SIGENERGY Energy Informatics Review, Vol.~6, Issue~2.}

\author{Boris Ruf}
\email{boris.ruf@axa.com}
\affiliation{%
  \institution{AXA AI Research}
  \city{Paris}
  \country{France}
}

\author{Marcin Detyniecki}
\email{marcin.detyniecki@axa.com}
\affiliation{%
  \institution{AXA AI Research}
  \city{Paris}
  \country{France}
}

\begin{abstract}
The rapid adoption of Retrieval-Augmented Generation (RAG) and long-context LLMs has shifted the computational burden toward the prefill phase---the initial stage where the model processes all input tokens in parallel before generating a response---yet its energy impact remains poorly characterised. We investigate the contradiction between massive hardware-level power spikes and the sub-linear marginal cost of input tokens reported in system-level studies. Through high-resolution telemetry of Llama-3.1 and Qwen-2 models on NVIDIA A100 hardware, we identify the efficiency threshold where $O(n^2)$ attention complexity overcomes GPU parallelisation. Our results quantify how optimised memory orchestration (PagedAttention) delays this scaling bottleneck across three orders of context scaling, yielding up to an 8.9$\times$ energy efficiency advantage at 112,000 tokens compared to standard baselines. Furthermore, we characterise the energy disparity between input and output tokens, demonstrating that optimised serving (vLLM) causes this ratio to narrow (from $\approx$850$\times$ to 750$\times$) as context scales due to high decode efficiency, whereas unoptimised baselines show a much smaller, roughly constant gap ($\approx$150--230$\times$) because their prefill is itself expensive. This work establishes context-scaling laws that differentiate engine-level optimisations from inherent model constraints, providing an empirical framework for sustainable LLM deployment.
\end{abstract}



\maketitle

\section{Introduction}
\label{sec:intro}
The rapid proliferation of energy-intensive Large Language Models (LLMs) has made efficiency a key focus of systems research. Frontier models are predominantly accessible via APIs that prevent direct energy measurement, resulting in reliance on approximations \cite{luccioni2024power}. Common estimation methodologies are based on output token counts, assuming that the primary driver of inference costs is autoregressive, sequential token generation, while input tokens are omitted \cite{ecologits}.

In practice, however, the vast majority of modern LLM applications are heavily dependent on massive input contexts. Retrieval-Augmented Generation (RAG), long-document summarisation, and complex coding assistants all represent use cases where the input volume significantly outweighs the generated output. 

This raises the question of how input tokens actually influence the energy balance, as recent research presents diverging results. Hardware-centric benchmarks like TokenPowerBench \cite{niu2025tokenpowerbench} report that inference engines such as vLLM \cite{kwon2023vllm}---a high-throughput LLM serving framework---spike GPU power draw by up to $3\times$ during prefill. This aligns with the $O(n^2)$ complexity of self-attention \cite{vaswani2017attention}, which arises because the mechanism requires pairwise comparisons between all tokens in the sequence, causing computational requirements to grow with the square of the context length. Conversely, systematic usage studies \cite{caravaca2025prompts} observe that processing even very large contexts (e.g., 50,000 tokens) consumes only marginally more total energy than small prompts, suggesting a sub-linear marginal cost. This creates an apparent ``Prefill Paradox'' where high instantaneous power draw is seemingly decoupled from total energy expenditure.

This discrepancy highlights a lack of understanding of how context-length scaling affects energy costs in LLM inference, emphasising the need to characterise the prefill and decode phases independently.

In this short paper, we address the following research questions:
\begin{enumerate}
    \item At what context length threshold does the $O(n^2)$ complexity of attention overcome GPU parallelisation, and how does this point of compute saturation differ across engines?
    \item How do distinct memory-orchestration strategies (PagedAttention vs. SDPA) influence the location and stability of the efficiency threshold as context length scales?
    \item How does the energy intensity disparity between input and output tokens evolve across varying context depths, and how is this relationship modulated by engine-level optimisations?
\end{enumerate}

\section{Background}
\label{sec:background}
LLM inference proceeds in two distinct phases with fundamentally different computational profiles \cite{patel2023splitwise}. The \textit{prefill} phase processes all input tokens in parallel: the model reads the entire prompt and computes a Key-Value (KV) cache that encodes the contextual relationships between tokens via the self-attention mechanism \cite{vaswani2017attention}. Because all tokens are processed simultaneously, this phase is \textit{compute-bound}---it fully utilises the GPU's parallel arithmetic units. The \textit{decode} phase then generates output tokens one at a time, where each new token attends to the full KV-cache built during prefill. This sequential dependency makes decoding \textit{memory-bound}---the GPU spends most of its time loading model weights and cache entries rather than performing arithmetic \cite{kwon2023vllm}.

This asymmetry has direct energy implications: prefill draws high instantaneous power (the GPU is fully active) but completes quickly, while decode draws lower power but runs for many sequential steps. Understanding how this balance shifts with context length is central to this work.

\section{Related Work}
\label{sec:related}

Understanding LLM energy consumption requires bridging hardware telemetry with application-level metrics. TokenPowerBench \cite{niu2025tokenpowerbench} provides a foundation for measuring instantaneous power draw by isolating prefill and decode stages, revealing that energy per token rises faster than parameter count due to memory traffic penalties. Caravaca et al.\ \cite{caravaca2025prompts} offer a longitudinal perspective, observing that while both input and output tokens contribute to energy footprints, output generation is significantly more intensive (e.g., a reported 11:1 energy disparity for output-heavy prompt configurations). This aligns with findings from the MELODI framework \cite{husom2024price} that energy consumption is weakly correlated with prompt length (0.075) but strongly correlated with response length (0.846). Complementing these energy-centric studies, LLMCO2 \cite{fu2025llmco2} improves inference-time carbon-footprint prediction, helping connect measured energy behavior to emissions-aware reporting.

Recent modelling efforts have quantified these dynamics using polynomial formulations. Wilkins et al.\ \cite{wilkins2024offline} and Wu et al.\ \cite{wu2025sustainable} model inference energy as a function of $T_{in}$ and $T_{out}$, highlighting a massive disparity where processing context for "memory formation" (e.g., RAG) can consume $1,000\times$ more energy than simple indexing. Nguyen et al.\ \cite{nguyen2024sustainable} further discuss system-level trade-offs in sustainable LLM serving, reinforcing that phase-specific efficiency must be evaluated together with deployment-level constraints. Furthermore, large-scale studies have diagnosed the underlying mechanisms of consumption, identifying system-wide bottlenecks such as memory pressure and KV-cache availability \cite{mlenergy2026joules}. In multimodal regimes, the prefill cost for non-text inputs (images/video) can increase by up to $15\times$ compared to text-only baselines \cite{mlenergy2026joules}.

Optimisation techniques attempt to mitigate these costs through structural interventions. These include memory management through PagedAttention (\textit{vLLM} \cite{kwon2023vllm}), phase-splitting (\textit{Splitwise} \cite{patel2023splitwise}), chunked prefills (\textit{Sarathi-Serve} \cite{agrawal2024sarathi}), and KV-cache materialization in flash storage (\textit{MatKV} \cite{shin2025matkv}). Additionally, the resource asymmetry between the prefill and decode phases has prompted new scheduling paradigms (\textit{DistServe} \cite{zhong2024distserve}) to address the "Sustainable AI Trilemma" \cite{wu2025sustainable} of balancing AI capability, digital equity, and environmental sustainability. In the same spirit, Sinha et al.\ \cite{sinha2025utilization} caution against relying on utilization as a standalone sustainability proxy and emphasize mechanism-level attribution for inference-serving decisions.

\section{Theoretical Analysis}
\label{sec:theory}

We hypothesise that the quadratic complexity of self-attention eventually dominates energy consumption due to three fundamental physical hardware limits:

\begin{enumerate}
    \item \textbf{High Parallel Efficiency Phase:} For small contexts, GPUs maximise throughput by processing tokens in parallel. In this regime, the energy per token effectively decreases as the fixed startup and weight-loading costs are amortised over more input tokens.
    \item \textbf{The Compute Saturation Cliff:} Once the context size $N$ exceeds the GPU's capacity for simultaneous parallel execution (limited by active hardware compute units), the engine must resort to sequential processing of attention chunks. At this point, the $O(n^2)$ complexity is no longer masked, and the energy tax becomes measurable.
    \item \textbf{The Memory Bandwidth Wall:} Calculating self-attention for massive sequences requires storing and frequently accessing the $N \times N$ attention probability matrix. As $N$ grows, the volume of data movement between the VRAM and compute cores scales quadratically, eventually saturating the memory bus. Energy consumption then becomes dominated by the high Joule-per-bit cost of off-chip data transfers.
\end{enumerate}

\begin{figure*}[t]
    \centering
    \begin{subfigure}{0.48\textwidth}
        \centering
        \includegraphics[width=\linewidth]{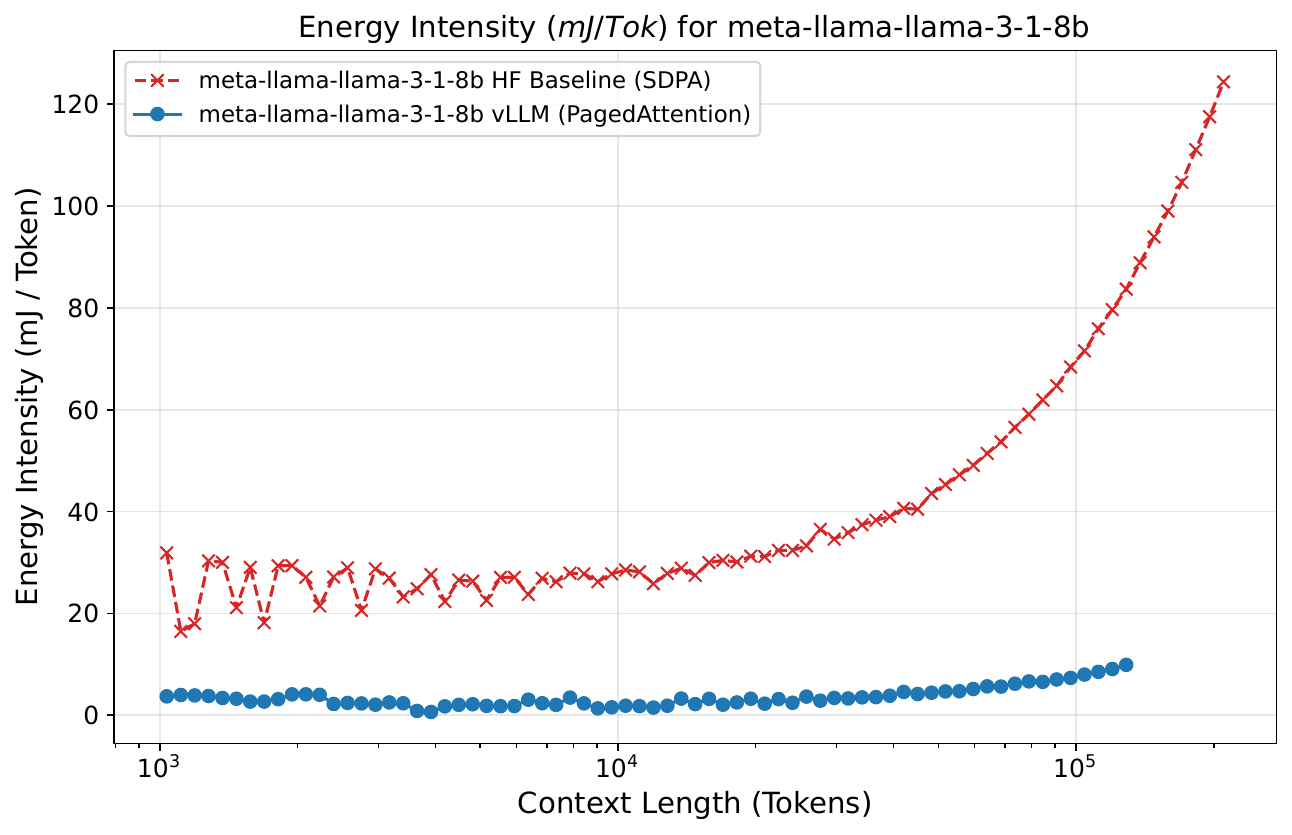}
        \caption{Llama-3.1 8B}
        \label{fig:intensity-llama}
    \end{subfigure}
    \hfill
    \begin{subfigure}{0.48\textwidth}
        \centering
        \includegraphics[width=\linewidth]{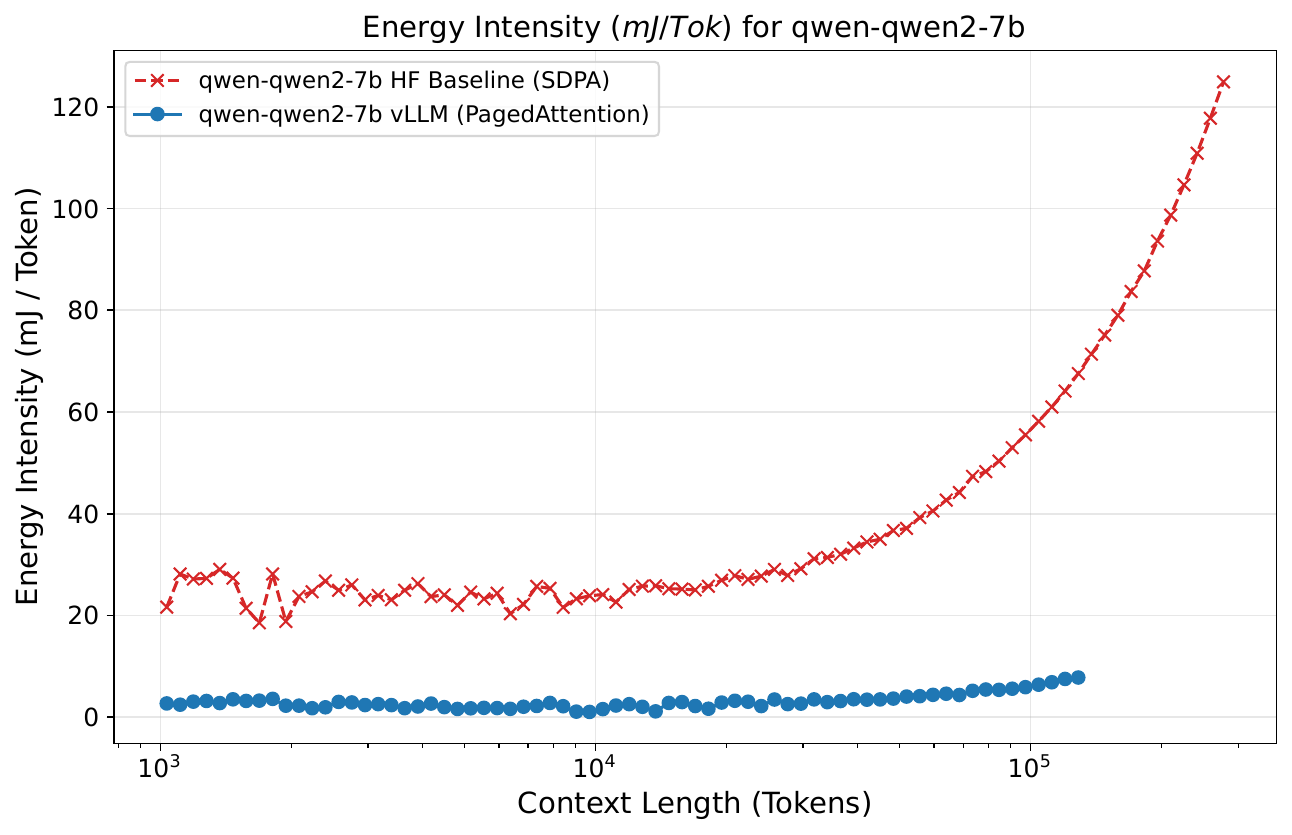}
        \caption{Qwen-2 7B}
        \label{fig:intensity-qwen}
    \end{subfigure}
    \caption{Energy Intensity ($mJ/Tok$) across context lengths. Note the efficiency threshold in the HF baseline compared to the delayed efficiency inflection point in vLLM.}
    \label{fig:intensity}
\end{figure*}

\section{Experimental Design}
\label{sec:design}
To investigate the energy impact of massive inputs, we implement a modular benchmarking suite designed to bridge the gap between coarse application-level estimates and fine-grained hardware telemetry. Our experimental design draws inspiration from the architecture of TokenPowerBench \cite{niu2025tokenpowerbench} and the feature-aware analysis of MELODI \cite{husom2024price}.

\subsection{Instrumentation and Telemetry}
We utilise a multi-threaded telemetry manager that samples GPU power draw via the NVIDIA Management Library (NVML) at a high-resolution 1ms sampling rate, consistent with NVML's documented minimum update interval for instantaneous power queries. While CPU/DRAM telemetry (e.g., RAPL) is often restricted in virtualised cloud environments, our methodology follows the hardware-centric approach of Caravaca et al.\ \cite{caravaca2025prompts}, focusing on the GPU as the primary energy consumer in LLM prefill phases. Each power sample is timestamp-aligned with the inference engine's event markers (prefill\_start and prefill\_end). In our benchmarks, the latter denotes the completion of the first-token generation, effectively isolating the context-processing phase. This allows us to integrate power over precisely defined intervals to calculate $E_{\text{prefill}}$.

\subsection{Workload Configuration}
To identify the exact efficiency threshold where $O(n^2)$ complexity breaks hardware parallelisation, we utilise a logarithmic context sweep. This approach provides the deterministic control necessary for high-resolution phase transition detection.

We generate synthetic prompts of length $L \in [2^8, 2^{17}]$ by repeating high-entropy base text and truncating to exact token counts. Since the attention mechanism performs identical matrix operations regardless of token content (i.e., the computational cost depends on sequence length, not semantic meaning), the use of repeated text does not affect the measured energy profile. We use model-specific tokenisers (e.g., Llama-3 128k and Qwen-2 1M) to ensure exact sequence lengths. Note that the maximum tested context length is bounded by each engine's supported context window; vLLM enforces the model's declared context limit (128k for Llama-3.1), whereas the HF Baseline permits extrapolation beyond this boundary, resulting in additional data points at higher context lengths in Figures~\ref{fig:intensity} and~\ref{fig:scaling}. This granular control eliminates the noise of variable semantic content and ensures that the $x$-axis (Context Length) is perfectly uniform, allowing for the precise calculation of the discrete derivative of energy consumption $\frac{dE}{dL}$. The benchmarks are conducted on NVIDIA A100 (80GB) instances across two model families (Llama-3.1 8B, Qwen-2 7B) using bfloat16 precision to account for architectural variability in attention implementation. This configuration (A100 hardware, bfloat16 precision, model-native tokenisers, NVML-based telemetry) is consistent with the experimental setups used in TokenPowerBench \cite{niu2025tokenpowerbench} and MELODI \cite{husom2024price}.

\subsection{Controlled Inference Scenarios}
We utilise two distinct inference engines as our experimental groups: (1) \textbf{Hugging Face Transformers} (hereafter \textbf{HF Baseline}), utilising the standard ``Eager'' implementation with Scaled Dot-Product Attention (SDPA) and contiguous KV-cache allocation; and (2) \textbf{vLLM (Optimised)} \cite{kwon2023vllm}, which uses PagedAttention to optimise memory management and minimise fragmentation. In both cases, we isolate the prefill phase by setting the generation length to exactly 1 token. Each configuration is preceded by a warm-up pass to stabilise GPU clock speeds and power states. As our workload is fully deterministic (fixed synthetic inputs, greedy decoding, no batching), each context length is measured in a single run; prior work on similar deterministic GPU benchmarks has shown negligible run-to-run variance under these conditions \cite{niu2025tokenpowerbench}.

\subsection{Comparative Intensity Protocol}
To characterise the energy disparity between input and output tokens, we conduct a secondary benchmarking run. In this configuration, we fix the output length to $G = 128$ tokens across the same logarithmic context sweep. By calculating the total energy $E_{total}$ and subtracting the isolated prefill energy $E_{prefill}$ obtained in the previous step, we derive the mean energy per output token $E_{out} = (E_{total} - E_{prefill}) / G$. This allows for a direct intensity comparison ($E_{out} / E_{in}$) across varying context depths.

\section{Experimental Evaluation}
\label{sec:evaluation}
Our evaluation characterises the energy-to-context scaling law across two model families (Llama-3.1 8B and Qwen-2 7B) on NVIDIA A100 (80GB) hardware. We analyse the resulting dataset through three primary lenses: (1) \textbf{Energy Intensity Scaling} to identify the hardware parallelisation threshold, (2) \textbf{Input vs. Output Proportion} to quantify energy disparity ($E_{out} / E_{in}$), and (3) \textbf{Component-Wise Attribution} to decompose the footprint using the correlation techniques introduced in MELODI \cite{husom2024price}. By comparing the standard Hugging Face SDPA baseline against the optimised vLLM engine (PagedAttention), we observe a dramatic divergence in energy intensity as context length scales.

\subsection{Energy Intensity Scaling}
The experimental data reveals that while both engines exhibit quadratic scaling tendencies at extreme context lengths, the marginal cost of an input token ($\frac{dE}{dL}$) begins to rise significantly earlier in the baseline implementation. This leads to an efficiency disparity between optimised and standard serving that is nearly an order of magnitude (see Figure \ref{fig:intensity}). As detailed in Table \ref{tab:results}, at approximately 112,000 tokens, the vLLM engine processed the Llama-3.1 input in 3.54 seconds (1,005 Joules), while the baseline required 25.54 seconds (7,600 Joules). This represents a 7.6$\times$ improvement in energy efficiency for the optimised engine (reaching up to 8.9$\times$ for Qwen-2). These results confirm Hypothesis~1 (Section~\ref{sec:theory}): at small contexts, both engines benefit from high parallel efficiency, but the threshold where Hypothesis~2 manifests differs by an order of magnitude between engines.

\begin{table}[h]
\centering
\small
\caption{Prefill Energy Intensity at $L \approx 112k$ tokens.}
\label{tab:results}
\begin{tabular}{lrrr}
\toprule
\textbf{Model / Engine} & \textbf{Lat. (s)} & \textbf{Energy (J)} & \textbf{J/Tok} \\
\midrule
\textit{Llama-3.1 8B} & & & \\
HF Baseline & 25.54 & 7,600.2 & 0.0678 \\
vLLM (Paged) & \textbf{3.54} & \textbf{1,005.1} & \textbf{0.0090} \\
\midrule
\textit{Qwen-2 7B} & & & \\
HF Baseline & 23.04 & 6,832.0 & 0.0610 \\
vLLM (Paged) & \textbf{2.71} & \textbf{769.6} & \textbf{0.0069} \\
\bottomrule
\end{tabular}
\end{table}

\subsection{Input vs. Output Proportion}
\begin{figure}[t]
    \centering
    \includegraphics[width=\linewidth]{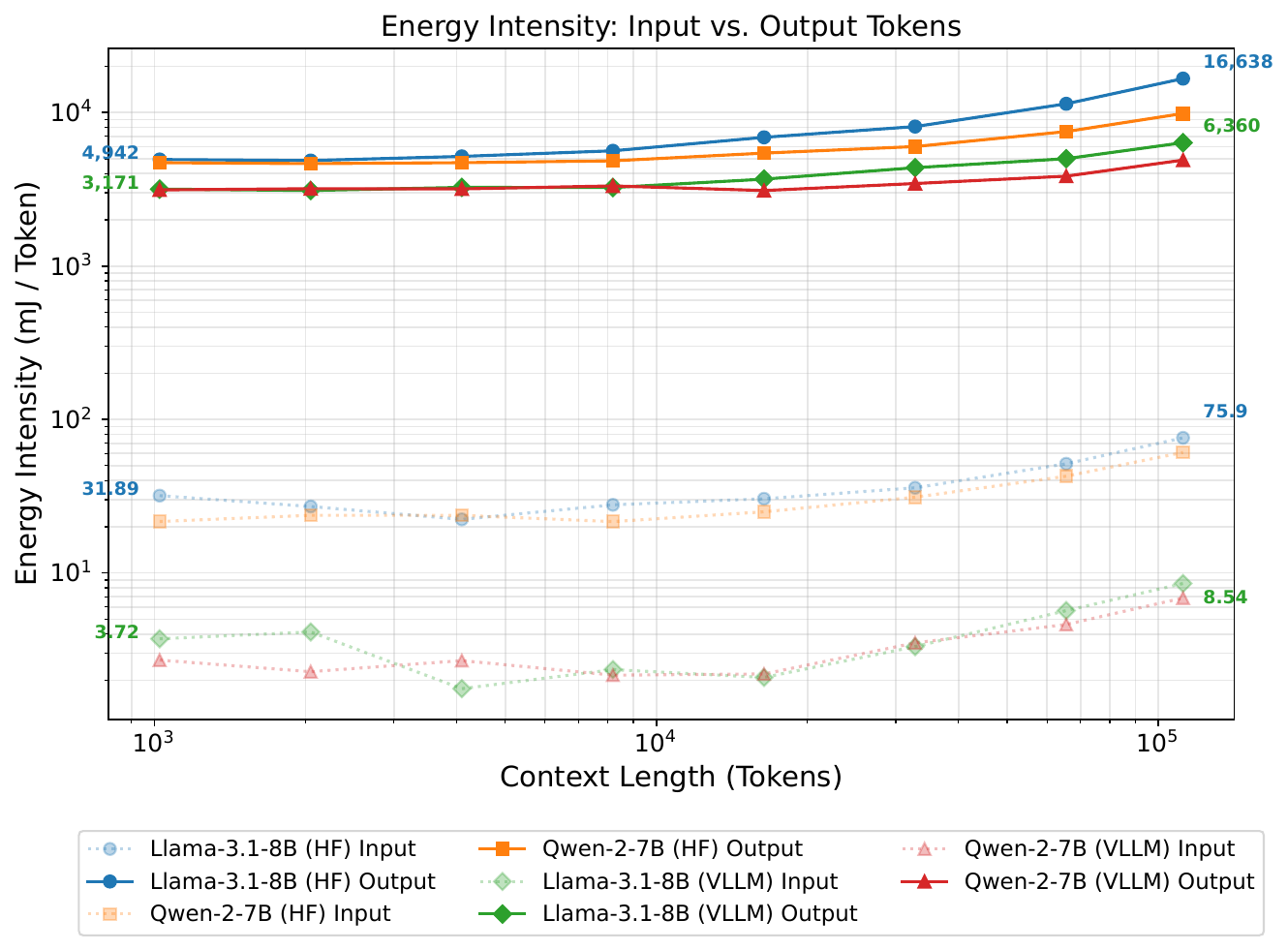}
    \caption{Comparative energy intensities ($mJ/Tok$) for input ($E_{in}$) and output ($E_{out}$) tokens. For optimised engines, the input intensity remains flat until $L > 32k$, while unoptimised engines pay a far higher $E_{in}$ throughout, which narrows their $E_{out}/E_{in}$ gap.}
    \label{fig:intensities}
\end{figure}

To provide a complete picture of operational costs, we analyse the energy relationship between prefill (input) tokens and decode (output) tokens (see Figure \ref{fig:intensities}). Output token energy is dominant in both regimes due to sequential dependencies, but the \textit{magnitude} of this dominance varies dramatically between optimised and unoptimised engines.

At small context lengths ($L=1k$), we observe that one output token consumes approximately 850$\times$ more energy than an input token in vLLM. Interestingly, as context scales to 112,000 tokens, this intensity disparity in optimised engines actually \textit{decreases} to $\approx$750$\times$. This is because vLLM's PagedAttention keeps the marginal cost of decoding low, while the $O(n^2)$ prefill math eventually increases the relative cost of input tokens.

In contrast, the HF Baseline ratio is far smaller and roughly flat, ranging from $\approx$155$\times$ at 1k to $\approx$220$\times$ at 112k context for Llama-3.1 (and from $\approx$220$\times$ to $\approx$160$\times$ for Qwen-2). The smaller gap is not a sign of efficiency: the baseline's prefill costs $\approx$9$\times$ more per token than vLLM's, while its decode costs only 1.5--2.6$\times$ more, so the expensive input tokens close much of the distance to the output tokens.

\subsection{Component-Wise Attribution}
\begin{figure}[t]
    \centering
    \includegraphics[width=\linewidth]{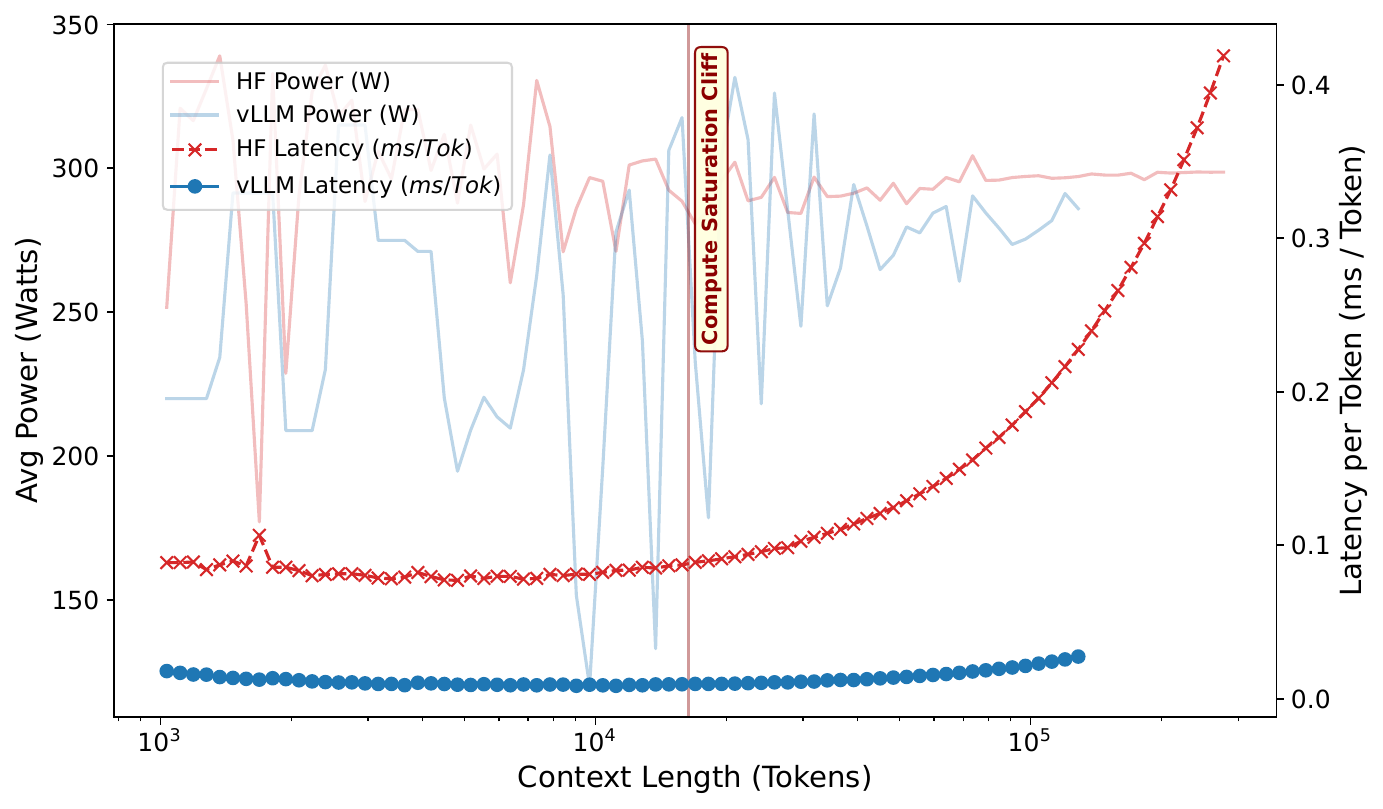}
    \caption{The Compute Saturation Cliff: Phase Map of Power vs. Normalised Latency. While power draw saturates early at ~310W, the latency per token begins to increase for unoptimised attention once hardware parallelisation is exceeded.}
    \label{fig:scaling}
\end{figure}

To investigate the drivers of the efficiency threshold, we analyse the relationship between instantaneous power draw ($P_{avg}$) and normalised latency (ms per token). Our telemetry indicates that for contexts exceeding 16,384 tokens, the GPU power draw plateaus at approximately 280W--310W, representing the thermal and electrical saturation of the A100—marking the entry into the \textbf{Compute Saturation Cliff} (see Figure \ref{fig:scaling}).

In this saturation regime, the energy intensity is primarily driven by \textit{latency extension} rather than increased power density. For the HF Baseline, we observe that normalised latency scales linearly ($O(n)$) with context length, implying that the total energy footprint is increasingly dominated by the quadratic ($O(n^2)$) total complexity of the \textbf{Memory Bandwidth Wall}. Specifically, the movement of massive attention matrices between VRAM and the compute cores accounts for the majority of the Joule-per-token increase. 

In contrast, vLLM's use of kernel fusion (FlashAttention \cite{dao2022flashattention}) and efficient memory orchestration (PagedAttention \cite{kwon2023vllm}) significantly reduces memory traffic by keeping intermediate attention states in on-chip SRAM. Our results show that vLLM's normalised latency remains nearly flat until $L > 32k$, implying that it manages to sustain \textbf{High Parallel Efficiency} much longer than the baseline. This attribution confirms all three hypotheses from Section~\ref{sec:theory}: the power plateau marks the Compute Saturation Cliff (Hypothesis~2), the linear latency growth reflects the Memory Bandwidth Wall (Hypothesis~3), and vLLM's ability to delay both transitions demonstrates sustained High Parallel Efficiency (Hypothesis~1).

\section{Discussion}
\label{sec:discussion}
The empirical profiling of LLM prefill energy reveals that the apparent contradiction in prefill energy scaling is primarily an orchestration challenge rather than an inherent arithmetic limitation. Unoptimised baselines hit an efficiency threshold because they fail to mask the quadratic growth of self-attention memory traffic. In contrast, optimised engines like vLLM maintain a flat $J/T$ profile across three orders of magnitude by aggressively amortising fixed hardware costs through parallel execution. This identifies the ``Prefill Paradox'' not as a contradiction in data, but as a manifestation of the ``Race to Finish'' principle: the high instantaneous power draw reported by hardware benchmarks is the precise mechanism that enables the sub-linear energy-per-token costs observed in system-level studies.

This divergence is explained by the fundamental difference between \textit{arithmetic intensity} and \textit{peak hardware utilisation}. Prefill's high arithmetic intensity enables the GPU to reach a state of maximum parallel efficiency, where power consumption per token is minimised despite high instantaneous power draw. The key implication is that, for modern optimised text-only stacks, input tokens remain effectively negligible relative to output tokens even at context lengths exceeding 100,000---though this balance may shift for multimodal inputs or future architectural changes.

\section{Conclusion}
\label{sec:conclusion}
By investigating the prefill energy contradiction, this research bridges the gap between hardware-level power measurements and application-level energy footprints. Our empirical results find that while GPU power draw saturates at $\approx$16k tokens, unoptimised baselines suffer from the $O(n^2)$ energy tax almost immediately once parallel overhead is exceeded. In contrast, optimised engines maintain an efficient linear $J/T$ profile until $L > 32k$. Memory-orchestration strategies like PagedAttention and FlashAttention prove to be critical factors, delaying the efficiency threshold and yielding up to an 8.9$\times$ energy efficiency advantage at 112k tokens. Furthermore, we demonstrate how engine optimisations shift the intensity disparity between input and output tokens: optimised systems show a narrowing gap as they manage memory effectively, while unoptimised systems exhibit a much smaller, roughly constant gap of $\approx$150--230:1 because their prefill is itself expensive. This explains why the power spike of prefill is a signature of parallel efficiency, shifting the focus of sustainable AI from model constraints to infrastructure-level optimisations.

\section{Future Work}
\label{sec:futurework}
Building on these findings, we identify three critical vectors for future energy-aware LLM research. First, \textit{high-throughput amortisation} studies should investigate how request and continuous batching specifically compress the decode-phase energy disparity by further amortising weight-loading costs. Second, research into \textit{precision-energy trade-offs} is required to evaluate whether the memory traffic reduction of FP8 or INT4 quantisation outweighs the potential power overhead of specialised low-precision tensor units. Finally, \textit{architectural scalability} sweeps should be extended to Mixture-of-Experts (MoE) and 70B+ parameter models to determine if the scaling bottleneck shifts predictably with increased parameter count and activated memory footprints.

\bibliographystyle{ACM-Reference-Format}
\bibliography{references}

\end{document}